\documentstyle[preprint,prb,aps]{revtex}
\begin{document}
\draft
\title{Surface plasmon in 2D Anderson insulator with interactions}
\author{T. V. Shahbazyan and M. E. Raikh}
\address{Department of Physics, University of Utah, Salt Lake City, UT
84112}
\maketitle

\begin{abstract}
We study the effect of interactions on the zero-temperature a.c.
conductivity of 2D Anderson insulator at low frequencies. We show
that the enhancement of the real part of conductivity due to the
Coulomb correlations in the occupation numbers of localized states
results in the change of the sign of imaginary part within a certain
frequency range. As a result, the propagation of a surface plasmon in a
{\em localized} system becomes possible. We analize the dispersion law
of the plasmon for the two cases: unscreened Coulomb interactions and
the interactions screened by a gate electrode spaced by
some distance from the electron plane.
\end{abstract}
\narrowtext

\section{Introduction}

It is believed that in two-dimensions (2D) even small disorder localizes
all electron states.\cite{lee85} First hint for the absence of the
mobility threshold in 2D came from the calculation of the
weak localization correction,  $\delta\sigma(\omega)$, to the conductivity
at finite frequency $\omega$. It was shown\cite{gor79} that in the
limit $k_Fl\gg 1$ one has
$\delta\sigma(\omega)/\sigma_0={1\over k_F l}\ln|\omega\tau|$,
where $k_F$ is the Fermi momentum, $l$ is the mean free path, and
$\tau$ is the elastic scattering time; $\sigma_0=(e^2/h)k_Fl$ is the
Drude conductivity. The logarithmic singularity in
$\delta\sigma(\omega)$ indicates that the zero-temperature ($T=0$)
conductivity is metallic only if $\omega\gg \omega_0$, where

\begin{eqnarray}\label{1}
\omega_0\sim {1\over\tau}e^{-\pi k_F l}
\end{eqnarray}
is the characteristic frequency which marks the crossover to the
exponentially localized regime.\cite{lee85} From the
finite size, $L$, correction to the zero-frequency conductivity, which
is of the order of ${1\over k_F l}\ln (L/l)$, one can estimate the
localization length $\xi$ as\cite{lee85}

\begin{eqnarray}\label{2}
\xi=l\exp\Biggl({\pi \over 2}k_F l\Biggr).
\end{eqnarray}
Therefore, starting from metal, one cannot describe the $T=0$ behavior
of the conductivity at frequencies $\omega\sim\omega_0$. The adequate
language for this region would be the language of the localized
states. Within this language the conductivity originates from the
transitions between the localized states induced by an external a.c.
field.

Let us briefly remind the corresponding derivation of
$\sigma(\omega)$, carried out by Mott\cite{mott70} in the strongly
localized regime. The Hamiltonian for strongly localized electrons reads

\begin{eqnarray}\label{3}
{\cal H}=\sum_i \epsilon_i c_i^{\dagger}c_i+
\sum_{ij} I_{ij} (c_i^{\dagger}c_j+c_j^{\dagger}c_i),
\end{eqnarray}
where $\epsilon_i$ is the energy of a localized
state centered at ${\bf r}={\bf r}_i$ and $I_{ij}$ is the overlap
integral which falls off exponentially with distance:
$I_{ij}=I_0\exp(-|{\bf r}_i-{\bf r}_j|/a)$, where $a$ is the
size of the wave-function of the localized electron. The general
expression for the conductivity is given by the Kubo formula
%%%%%%%%%%%%%%%%%%%%%%%%%%%%%%%%
\begin{eqnarray}\label{4}
\sigma(\omega)={ie^2\omega\over \hbar A}
\sum_{\bf ij}|\langle {\bf i}|x|{\bf j} \rangle|^2
{n_{\bf i}-n_{\bf j} \over \omega + \omega_{\bf ij} + i0}
\end{eqnarray}
%%%%%%%%%%%%%%%%%%%%%%%%%%%%%%%%
where $\langle {\bf i}|x|{\bf j} \rangle$ is the matrix element of
$x$ calculated from the {\em exact}
eigenstates $|{\bf i} \rangle$ and $|{\bf j} \rangle$ of the
Hamiltonian (\ref{3}) with energies $E_{\bf i}$ and $E_{\bf j}$,
$\hbar\omega_{\bf ij}=E_{\bf i}-E_{\bf j}$ and $n_{\bf i}$ is the
occupation number of the state $|{\bf i} \rangle$.

The dissipative conductivity, $\mbox{Re}\sigma(\omega)$, is determined by
the pairs of states with $\omega_{\bf ij}=\omega$.
At $\hbar\omega\ll I_0$ the spatial separation between bare
states $|i\rangle$ and $|j\rangle$ is much larger than $a$. Then in the
calculation of eigenstates of a resonant pair one should take
into account the overlap $I_{ij}$ within this pair {\em only} and neglect
the overlap with all the other localized states. This gives
%%%%%%%%%%%%%%%%%%%%%%%%%%%%%%%%%
\begin{eqnarray}\label{5}
|{\bf i}\rangle = {\epsilon_i -\epsilon_j\over\Gamma}|i\rangle +
{2I_{ij}\over\Gamma}|j\rangle,~~~~~~
|{\bf j}\rangle = {2I_{ij}\over\Gamma}|i\rangle +
{\epsilon_j-\epsilon_i \over\Gamma}|j\rangle.
\end{eqnarray}
%%%%%%%%%%%%%%%%%%%%%%%%%%%%%%%%%
The corresponging energies are
%%%%%%%%%%%%%%%%%%%%%%%%%%%%%%
\begin{eqnarray}\label{6}
E_{\bf i,j}=(\epsilon_i+\epsilon_j)/2 \pm \Gamma/2,~~~~~
\Gamma=[(\epsilon_i-\epsilon_j)^2+4I_{ij}^2]^{1/2}=\hbar\omega_{\bf ij}.
\end{eqnarray}
%%%%%%%%%%%%%%%%%%%%%%%%%%%%%
Using (\ref{5}) the matrix element in (\ref{4}) takes the form

\begin{eqnarray}\label{7}
\langle {\bf i}|x|{\bf j}\rangle=(x_i-x_j)I_{ij}/\Gamma.
\end{eqnarray}
The contribution to $\mbox{Re}\sigma(\omega)$ from
pairs with shoulder ${\bf r}$ can be presented as

\begin{eqnarray}\label{8}
\mbox{Re}\sigma(\omega,{\bf r})=
{e^2\over\hbar}{\pi x^2 I^2(r)\over \hbar \omega}F(\hbar\omega,r),
\end{eqnarray}
where $F(\hbar\omega,r)$ is the density of pairs with shoulder
${\bf r}$ and
excitation energy $\hbar\omega$ [here $I(r)=I_0e^{-r/a}$]. The
density $F(\hbar\omega,r)$ is determined by the condition
that the pair is singly occupied (with energies on the opposite sides
from the Fermi level). Then we have

\begin{eqnarray}\label{9}
F(\hbar\omega,r)&=&g^2\int d\epsilon_1\int d\epsilon_2
\theta\Biggl({\hbar\omega\over 2}+{\epsilon_1+\epsilon_2\over 2}\Biggr)
\theta\Biggl({\hbar\omega\over 2}-{\epsilon_1+\epsilon_2\over 2}\Biggr)
\delta\Biggl[\sqrt{(\epsilon_1-\epsilon_2)^2+4I^2(r)}-\hbar\omega\Biggr]
\nonumber\\
&=&{2g^2(\hbar\omega)^2\over\sqrt{(\hbar\omega)^2-4I^2(r)}},
\end{eqnarray}
where $g$ is the density of states and $\theta(x)$ is the
step-function. Substituting (\ref{9}) into
(\ref{8}) and integrating over ${\bf r}$ we get

\begin{eqnarray}\label{10}
\mbox{Re}\sigma(\omega)=
{e^2\over\hbar}(2\pi^2 g^2 \hbar\omega)
\int_0^{\infty}{drr^3I^2(r)\over\sqrt{(\hbar\omega)^2-4I^2(r)}}.
\end{eqnarray}
For $\hbar\omega\ll I_0$ the main contribution to the integral comes
from $r\sim r_{\omega}=a\ln(2I_0/\hbar\omega)\gg a$ and one obtains

\begin{eqnarray}\label{11}
\mbox{Re}\sigma(\omega)=\sqrt{2}\pi^2
{e^2\over \hbar}(g^2a\hbar^2\omega^2 r_{\omega}^3).
\end{eqnarray}

Although we cannot show it explicitly, we argue below that the Mott
expression (\ref{11}) for $\mbox{Re}\sigma(\omega)$ is valid also
for the Anderson insulator with $k_Fl\gg 1$. When applying
(\ref{11}) to the Anderson insulator one should replace $a$ by $\xi$
from (\ref{2}) and substitute $g=m/2\pi\hbar^2$ ($m$ is the electron
mass). However, the question
remains: what is the magnitude of $I_0$ in this case? A plausible
estimate can be obtained in the spirit of the Thouless picture of
localization.\cite{tho77}
By definition, $I_0$ represents the splitting of energy
levels of two neighboring localized states (with centers at
distance $\sim \xi$). The estimate for $I_0$ emerges if one equates
this splitting to the mean energy spacing for localized
states centered within the area $\sim \xi^2$, so that $I_0\sim 1/g\xi^2$
(see also Ref.~\onlinecite{imr93}).

Note that in 1D case a similar argument leads to
$I_0\sim 1/g_1\xi$, $g_1$ being the 1D density of states.
Important is that in 1D the Kubo formula can be evaluated
exactly\cite{ber73} resulting in the 1D version of the Mott
formula, from which one can recover the above estimate for $I_0$.
Such a mapping was first established by Shklovskii and
Efros.\cite{shk81}

Using Eq.~(\ref{10}) one can formally evaluate
$\mbox{Re}\sigma(\omega)$ for $\hbar\omega > 2I_0$. In this case the
main contribution to the integral comes from $r\sim a$ and we get
%%%%%%%%%%%%%%%%%%%%%%%%%%%%%%%%
\begin{eqnarray}\label{12}
\mbox{Re}\sigma(\omega)={3\pi^2\over 4}
{e^2\over \hbar}(I_0^2g^2a^4).
\end{eqnarray}
%%%%%%%%%%%%%%%%%%%%%%%%%%%%%%%%
Certainly, the presentation of $I(r)$ in the form $I_0e^{-r/a}$ makes
sense only for $r\gg a$ This means that the numerical coefficient in
Eq.~(\ref{12}) is not reliable.

Clearly, at large frequencies $\omega\gg\omega_0$ the conductivity of
the Anderson insulator
should have the Drude form. The fact that Eq.~(\ref{12}), calculated
for strongly localized electrons, is also frequency independent allows
us to assume that the description of a.c. transport in the Anderson
insulator based on Hamiltonian (\ref{3}) is accurate within a
numerical coefficient. In other words, we assume that despite the
complex structure of the electron wave-functions in the Anderson
insulator the energy dependence of the matrix elements calculated
between these functions is still given by (\ref{7}).

Eq.~(\ref{12}) provides yet another way to estimate $I_0$ in the limit
$k_Fl\gg 1$. Namely, it matches $\sigma_0$ if we take
$I_0\sim (k_Fl)^{1/2}/g\xi^2$. We see that the dependence of $I_0$ on
$\xi$ in both estimates is the same; the extra factor $(k_Fl)^{1/2}$
presumably can be accounted for the $\ln(k_Fl)$ corrections to the
exponent of $\xi$.

There is also another argument in favor of the above estimate for
$I_0$. The frequency dependence of $\mbox{Re}\sigma(\omega)$ in the
Anderson insulator becomes strong for $\omega\ll\omega_0$, whereas in the
picture of strongly localized electrons the demarcation frequency is
$\omega\sim I_0/\hbar$. Equating $I_0$ to $\hbar\omega_0$ we get
$I_0=k_Fl/g\xi^2$, with another extra factor $k_Fl$.

The simplified description of the Anderson insulator based on the
Hamiltonian (\ref{3}) allows one to include into consideration
the Coulomb correlations (i.e., the correlations in the occupation
numbers of the localized states caused by electron-electron
interactions) using the ideas first spelled
out in Refs.~\onlinecite{efr81,shk81,efr85}. This is the main goal of the
present paper. We study the effect of Coulomb correlations on both
$\mbox{Re}\sigma(\omega)$ and $\mbox{Im}\sigma(\omega)$. The most
drastic conclusion we come to is that due to modification of
$\mbox{Im}\sigma(\omega)$ by the Coulomb correlations
{\em a system of localized
electrons can support surface plasmons within a certain frequency
range}. We also show that these plasmons cause an additional structure
in the behavior of $\mbox{Re}\sigma(\omega)$ at
$\omega > \omega_0 \sim I_0/\hbar$.

The paper is organized as follows. In Section \ref{sec2} we analyze
the polarizability of the localized system in the absence of Coulomb
correlations. In Section \ref{sec3} we introduce the Coulomb correlations
and find the dispersion law for the surface plasmons. In
Section \ref{sec3} we study the corrections to the dispersion law due
to the resonant scattering of plasmons by pairs of localized states.
In Section \ref{sec4} we calculate the plasmon contribution to
$\mbox{Re}\sigma(\omega)$. Section \ref{sec5} concludes the paper.

\section{Polarizability in the absence of Coulomb correlations}
\label{sec2}

In this section we demonstrate that without Coulomb correlations the
2D Anderson insulator cannot support surface plasmon.

In the framework of the linear response theory the dispersion law of a
plasmon, $\omega(q)$, is determined from the condition\cite{mahan}

\begin{eqnarray}\label{13}
1=v(q)\mbox{Re}{\bf P}(\omega,q),
\end{eqnarray}
where $v(q)$ is the Fourier component of the electron-electron
interaction $v(r)$ and ${\bf P}(\omega,q)$ is the polarization operator.
Within a standard approach ${\bf P}(\omega,q)$ is calculated for
non-interacting electrons described by Hamiltonian (\ref{3}):

\begin{eqnarray}\label{14}
P(\omega,q)={1\over A}
\sum_{\bf ij}|\langle {\bf i}|e^{i{\bf qr}}|{\bf j} \rangle|^2
{n_{\bf i}-n_{\bf j} \over \hbar\omega + E_{\bf i}-E_{\bf j} + i0}.
\end{eqnarray}
In the absence of disorder the eigenstates $|{\bf i} \rangle$ and
$|{\bf j} \rangle$ are the plain waves so that the matrix element in
(\ref{14}) reduces to the delta-function $\delta({\bf i-j-q})$. Then
evaluating $P(\omega,q)$ and substituting it into (\ref{13})
together with 2D Coulomb interaction $v(q)=2\pi e^2/\kappa q$
($\kappa$ is the dielectric constant) yields the surface
plasmon with the well-known dispersion law

\begin{eqnarray}\label{15}
\omega(q)=\Biggl[{2\pi n e^2 q\over m\kappa}\Biggr]^{1/2},
\end{eqnarray}
where $n$ is the 2D concentration of electrons.
The plasmon mode is undamped if $q < \omega/v_F$ where
$v_F=(4\pi n)^{1/2}\hbar/m$ is the Fermi velocity (for larger $q$
the Landau damping leads to a finite $\mbox{Im}P$). The latter
condition can be rewritten as $q<1/2a_B$ where
$a_B=\hbar^2\kappa/me^2$ is the Bohr radius.

In the case of a strong disorder the eigenstates ${\bf i}$ and ${\bf j}$
in (\ref{14}) are the localized states. The polarization operator
(\ref{14}) can be evaluated in a way similar to that employed in
the Introduction for calculation of a.c. conductivity. For small $q$
the matrix element in (\ref{14}) can be evaluated using (\ref{7}):

\begin{eqnarray}\label{16}
\langle {\bf i}|e^{i{\bf qr}}|{\bf j}\rangle=i{\bf qr}I(r)/\Gamma.
\end{eqnarray}
Then the contribution to $P(\omega,q)$  from the pairs
with the shoulder $r$ [density ${\cal P}(\omega,q,{\bf r})$ of
the polarization operator] can be presented as

\begin{eqnarray}\label{17}
\mbox{Re}{\cal P}(\omega,q,{\bf r})=({\bf qr})^2I^2(r)
\int_{2 I(r)}^{\infty}{d\Gamma\over \Gamma }
{F(\Gamma,r)\over (\hbar\omega)^2-\Gamma^2},
\end{eqnarray}
where $\Gamma$ and $F(\Gamma,r)$ is given by Eqs.~(\ref{6}) and
(\ref{9}), respectively (the integral is understood as principal
part). Substituting $F(\Gamma,r)$ into
(\ref{17}) and integrating over $\Gamma$ we obtain

\begin{eqnarray}\label{18}
\mbox{Re}{\cal P}(\omega,q,{\bf r})&=&2g^2({\bf qr})^2I^2(r)
\int_{2I(r)}^{\infty}d\Gamma
{\Gamma\over \sqrt{\Gamma^2-4I^2(r)}[(\hbar\omega)^2-\Gamma^2]}
\nonumber\\
&=&-{\pi g^2({\bf qr})^2I^2(r)\over \sqrt{4I^2(r)-(\hbar\omega)^2}},
{}~~~~~~~\mbox{for}~~\hbar\omega<2I(r),
\nonumber\\
&=&0,~~~~~~~~~~~~~~~~~~~~~~~~~~~~\mbox{for}~~\hbar\omega>2I(r).
\end{eqnarray}
We see that $\mbox{Re}{\cal P}(\omega,q,{\bf r})$ is either negative or
zero, so that Eq.~(\ref{13}) cannot be satisfied.

In fact, the result (\ref{18}) is almost obvious. Indeed, the
imaginary part of polarization operator density,
$\mbox{Im}{\cal P}(\omega,q,{\bf r})$,
at small $q$ differs from $\mbox{Re}\sigma(\omega,{\bf r})$ in
(\ref{8}) by a factor $e^2\omega/q^2$, so it follows from (\ref{8})
and (\ref{9}) that

\begin{eqnarray}\label{19}
\mbox{Im}{\cal P}(\omega,q,{\bf r})
&=&-{\pi g^2({\bf qr})^2I^2(r)\over \sqrt{(\hbar\omega)^2-4I^2(r)}},
{}~~~~~~~\mbox{for}~~\hbar\omega>2I(r),
\nonumber\\
&=&0,~~~~~~~~~~~~~~~~~~~~~~~~~~~~\mbox{for}~~\hbar\omega<2I(r).
\end{eqnarray}
Since $\mbox{Re}{\cal P}(\omega,q,{\bf r})$ and
$\mbox{Im}{\cal P}(\omega,q,{\bf r})$ are connected via the Kramers-Kronig
relation, the form (\ref{19}) immediately follows from (\ref{18}).

\section{Coulomb correlations and surface plasmon}
\label{sec3}

In the previous section the polarization operator was evaluated using
the pair density (\ref{9}) which was derived for non-interacting
electrons. As it was first pointed out by Efros\cite{efr81},
interactions modify strongly the density of singly-occupied pairs. The
underlying physics is the following. A pair can be singly-occupied
even if both energy states reside below the Fermi level. The right
condition for the pair to be singly-occupied is that the addition of a
second electron (which interacts with the first one) is energetically
unfavorable. Such a Coulomb correlations effectively enhance the
density of ``soft'' pairs (i.e., the pairs with small excitation
energy $\Gamma$).

Our goal is to apply the latter argument, which was presented for
strongly localized system,  to the Anderson insulator with large
$\xi$. In order to do so we will adopt two assumptions:

$(i)$ The interactions do not change the localization radius $\xi$.

$(ii)$ The estimate for the overlap integral, $I_0\sim 1/g\xi^2$,
is unchanged in the presence of interactions.

In other words, we assume that switching on the interactions leads to
the Coulomb shifts of the eigenenergies but does not affect the wave
functions. Note that assumptions $(i)$ and $(ii)$ contradict
those made in Refs.~\onlinecite{ale94} and \onlinecite{pol93},
respectively.

As we will see below, the relevant pairs would be those with shoulder
$r\sim\xi$. In other words, the relevant transitions shift the
position of electron by $\sim\xi$. To establish the form of the density
of singly occupied pairs $F(\Gamma,r)$ with such a shoulder one can
argue as follows. An isolated region of a size $\xi$ can be viewed as a
small metallic granule. The transfer of an additional electron into
this granule leads to the charging energy $U=e^2/2C$, where $C$ is the
capacitance of a granule. In other words, the levels in the granule
get shifted\cite{shk82} by an amount $\sim U$. Consider now two
neighboring granules and assume that $U\gg I_0$. Due to aforementioned
charging effect the highest occupied levels in the two granules
typically differ by $\sim U$. Let for concreteness the highest
occupied level in the first granule be higher by $U$ than in the
second one. Then the sought singly occupied pair with frequency
$\Gamma$ can be composed from the top occupied states in the first
granule (these states should belong to the energy interval $\Gamma+U$
measured from the highest occupied level) and the empty states in the
second granule. Then the density of pairs $F(\Gamma,r)$
(with $r\sim \xi$) can be estimated as $g^2(\Gamma+U)$. With the
energy splitting taken into account it can be written as

\begin{eqnarray}\label{20}
F(\Gamma,r)=
{2g^2\Gamma(\Gamma +U)\over\sqrt{\Gamma^2-4I^2(r)}}.
\end{eqnarray}
Then at $U=0$ we return to (\ref{9}). Certainly, our consideration,
based on artificial arranging the localized states into the granules,
provides only the
order of magnitude estimate of $F(\Gamma,r)$. In particular, the
numerical coefficient in (\ref{20}) cannot be found from such a
consideration. Our choice of numerical coefficient in (\ref{20})
provides matching with a similar expression for strongly localized
regime.\cite{shk81,efr85,efr85b}

With the pair density (\ref{20}) we can now easily evaluate
$\mbox{Re}{\cal P}(\omega,q,{\bf r})$. Calculating the integral over
$\Gamma$ in (\ref{17}) yields

\begin{eqnarray}\label{21}
\mbox{Re}{\cal P}(\omega,q,r)
&=&-{\pi g^2({\bf qr})^2I^2(r)\over \sqrt{4I^2(r)-(\hbar\omega)^2}}
\Biggl\{1+{2\over\pi}{U\over\hbar\omega}
\arctan\Biggl[{\hbar\omega\over\sqrt{4I^2(r)-(\hbar\omega)^2}}\Biggr]\Biggr\},
{}~~~\mbox{for}~~\hbar\omega<2I(r),
\nonumber\\
&=&{U\over\hbar\omega}
{2 g^2({\bf qr})^2I^2(r)\over \sqrt{(\hbar\omega)^2-4I^2(r)}}
\ln\Biggl[{\sqrt{(\hbar\omega)^2-4I^2(r)}+\hbar\omega\over2I(r)}\Biggr],
{}~~~~~~\mbox{for}~~\hbar\omega>2I(r).
\end{eqnarray}
The expression (\ref{21}) for $\mbox{Re}{\cal P}(\omega,q,r)$ can be
also obtained, using the Kramers-Kronig relations, from
$\mbox{Im}{\cal P}(\omega,q,r)$ which has a simple form

\begin{eqnarray}\label{22}
\mbox{Im}{\cal P}(\omega,q,{\bf r})=
-{\pi ({\bf qr})^2I^2(r)\over (\hbar\omega)^2}F(\hbar\omega,r)
&=&-{\pi g^2({\bf qr})^2I^2(r)\over \sqrt{(\hbar\omega)^2-4I^2(r)}}
\Biggl(1+{U\over\hbar\omega}\Biggr),~~~\mbox{for}~~\hbar\omega>2I(r),
\nonumber\\
&=& 0,~~~\mbox{for}~~\hbar\omega<2I(r).
\end{eqnarray}
Note that as in (\ref{19}),
$\mbox{Im}{\cal P}(\omega,q,{\bf r})\neq 0$ only for
$\hbar\omega>2I(r)$.

We see that the enhancement in the density of pairs with small
$\Gamma$ leads to a {\em positive} sign of $\mbox{Re}{\cal P}(\omega,q,r)$
for $\hbar\omega > 2I(r)$. This is our main observation.

In order to obtain $\mbox{Re}P(\omega,q)$
one should integrate $\mbox{Re}{\cal P}(\omega,q,r)$ over ${\bf r}$.
For $\hbar\omega > 2I_0$ the main contribution to this integral comes from
$r\sim\xi$ [like in derivation of Eq.~(\ref{12})] and we get

\begin{eqnarray}\label{23}
\mbox{Re}P(\omega,q)={3\pi\over 4}
{q^2UI_0^2g^2\xi^4\over (\hbar\omega)^2}
\ln\Biggl({\hbar\omega\over I_0}\Biggr).
\end{eqnarray}
For generality we will assume that there is also a gate at a distance $d$
from the plane of 2D electrons. In this case the Fourier component of
electron-electron interaction has the form

\begin{eqnarray}\label{24}
v(q)={2\pi e^2\over \kappa q}\Biggl(1-e^{-2qd}\Biggr).
\end{eqnarray}
If the gate is close to the electron plane, that is $qd\ll 1$, we
can expand the exponent in (\ref{24}) and get $v(q)=4\pi e^2d/\kappa$.
If $d\ll \xi$ then the capacitance $C$ reduces to the capacitance of
two disks with  area $S=\xi^2$ separated by a distance $d$, so that
$C=\kappa S/4\pi d$ and consequently
$U=4\pi e^2d/\kappa\xi^2$. With these $v(q)$ and $U$, after
substituting (\ref{23}) into the plasmon equation (\ref{13}), we obtain

\begin{eqnarray}\label{25}
1={3\pi\over 4}(q\xi)^2\Biggl({U\over\hbar\omega}\Biggr)^2
(I_0g\xi^2)^2\ln\Biggl({\hbar\omega\over I_0}\Biggr).
\end{eqnarray}
Certainly, the numerical coefficient in (\ref{25}) should not be taken
seriously. According to the assumtion $(ii)$, $I_0\sim g\xi^2$. Then
Eq.~(\ref{25}) yields the following dispersion law for the surface
plasmon

\begin{eqnarray}\label{26}
q(\omega)={1\over \xi}\Biggl({\hbar\omega\over U}\Biggr)
\ln^{-1/2}\Biggl({\hbar\omega\over I_0}\Biggr).
\end{eqnarray}
We see that the dispersion law is close to acoustic.

Let us establish the frequency range for the surface plasmon with
dispersion law (\ref{26}). The validity of expansion (\ref{16}),
$q\xi\ll 1$, implies that
$U\gg \hbar\omega\ln^{-1/2}(\hbar\omega/I_0)$. On the other hand,
$\hbar\omega>2I_0$. Then the frequency range for plasmon is
$2I_0\lesssim\hbar\omega\lesssim U$. The nesessary condition for this
range to be wide is $U\gg I_0$. The ratio $U/I_0=4\pi e^2d/\xi^2 I_0$
with $I_0\sim 1/g\xi^2$ can be presented as $8\pi^2 d/a_B$, where
$a_B=\hbar^2\kappa/me^2$ is the Bohr raduis. Thus, the condition $d\gg a_B$
insures that the plasmon equation (\ref{13}) has a solution within a
wide frequency range. If $d<a_B$, the screening of Coulomb interaction
by the gate is strong and the number of soft pairs is not sufficient to
change the sign of $\mbox{Re}P(\omega,q)$.

A similar condition can
be obtained from the analysis of $\mbox{Im}P(\omega,q)$, which
can be derived by integration of Eq.~(\ref{22}) over ${\bf r}$

\begin{eqnarray}\label{27}
\mbox{Im}P(\omega,q)={3\pi^2\over 8}
{q^2I_0^2g^2\xi^4\over (\hbar\omega)^2}(\hbar\omega+U)
\sim {q^2\over (\hbar\omega)^2}(\hbar\omega+U).
\end{eqnarray}
The origin of $\mbox{Im}P(\omega,q)$ is the interaction of a
plasmon with ``resonant'' pairs having excitation energy $\omega$. In
fact, $\mbox{Im}P(\omega,q)$ describes the resonant scattering of a
plasmon by a pair of localized states. One can introduce a mean free
path, $l$, associated with such a scattering and obtain
$ql\sim\ln^{1/2}(\hbar\omega/I_0)/(1+\hbar\omega/U)$.
Thus, the condition $2I_0\lesssim\hbar\omega\lesssim U$ reduces to the
condition $ql\gtrsim 1$.

In the absense of gate [$qd\gg 1$ in Eq.~(\ref{24})] one should take
$U=e^2/\kappa\xi$, so that $U/I_0\sim \xi/a_B\gg 1$. Then after a
simple algebra we obtain

\begin{eqnarray}\label{28}
q(\omega)={1\over \xi}\Biggl({\hbar\omega\over U}\Biggr)^2
\ln^{-1}\Biggl({\hbar\omega\over I_0}\Biggr).
\end{eqnarray}
It can be shown that the above analysis of the validity applies in
this case as well and leads to the same frequency range
$2I_0\lesssim\hbar\omega\lesssim U$. Within this range we again have
$q\xi\ll 1$.

Thus, one should use Eq.~(\ref{26}) for $qd<1$ and Eq.~(\ref{28}) for
$qd>1$. On the other hand, the magnitude of $U$ depends on the ratio
$d/\xi$. Note that for $d>\xi$ one still can have $qd<1$. In this case
the dispersion law is given by Eq.~(\ref{26}) with $U=e^2/\kappa\xi$.

\section{Renormalization of the plasmon dispersion law}
\label{sec4}

In the previous section, when calculating the polarization operator,
we took into account the Coulomb correlations within a pair, but
neglected the effect of polarization of surrounding pairs on a given
pair. On the other hand, by averaging of the polarization operator
(\ref{14}) over frequencies of pairs $\Gamma$ and their shoulders
${\bf r}$ we have effectively replaced the localized system by a
medium. The average polarization of this medium gave rise to a plasmon
mode. Within this procedure the ``feedback'' from surrounding pairs
reduces to the interaction of a given pair with plasmons. In the present
section we study the renormalization of the plasmon spectrum due to
this effect.

Generally, the plasmon excitation is defined as a pole in the
density-density correlation function, $\Pi(\omega, {\bf q},{\bf q}')$,
which is related to the polarization operator
${\bf P}(\omega, {\bf q},{\bf q}')$ by the Dyson equation

\begin{eqnarray}\label{29}
\Pi(\omega, {\bf q},{\bf q}')=
{\bf P}(\omega, {\bf q},{\bf q}')+\int {d{\bf q}_1\over (2\pi)^2}
{\bf P}(\omega, {\bf q},{\bf q}_1)v(q_1)\Pi(\omega, {\bf q}_1,{\bf q}').
\end{eqnarray}
Before averaging, both ${\bf P}(\omega, {\bf q},{\bf q}')$ and
$\Pi(\omega, {\bf q},{\bf q}')$ depend on two momentum variables
${\bf q}$ and ${\bf q}'$. The approximation we made above reduces to
replacing of ${\bf P}(\omega, {\bf q},{\bf q}')$ by its average
${\bf P}(\omega, q)$, so that the solution of (\ref{29}) takes
the form

\begin{eqnarray}\label{30}
\Pi(\omega, q)=
{{\bf P}(\omega, q)
\over 1-v(q){\bf P}(\omega, q)}.
\end{eqnarray}
Then the  pole of $\Pi(\omega,q)$ is determined by the plasmon equation
(\ref{13}).

As a next step, we took for ${\bf P}(\omega, q)$ its expression
(\ref{14}) for non-interacting electrons, which represents a sum of
polarizations of pairs,

\begin{eqnarray}\label{31}
P(\omega,q)={1\over A}
\sum_{\bf ij}|\langle {\bf i}|e^{i{\bf qr}}|{\bf j} \rangle|^2
P_{\bf ij}(\omega),~~~~~
P_{\bf ij}(\omega)=
{n_{\bf i}-n_{\bf j} \over \hbar\omega + E_{\bf i}-E_{\bf j} + i0},
\end{eqnarray}
and performed the summation neglecting  correlations between the pairs
but with the Coulomb correlations within a pair included.

The renormalized ${\bf P}_{\bf ij}(\omega)$ for a given pair
can be obtained from the following procedure.
The function $\Pi(\omega, q)$ has a diagrammatic presentation in
a form of a series of bubbles $({\bf ij})$, corresponding to
$P_{\bf ij}$, connected by the
Coulomb interaction lines (see Fig.~1a). First, we arrange into a single
block the sum over all combinations of bubbles which appear between two
bubbles $({\bf ij})$ (see Fig.~1b).
Then we replace this block by its average, so that the result
can be presented as two bubbles $({\bf ij})$ connected by
a plasmon propagator $\Pi(\omega, q)$ (see Fig.~1b) (there is
also an extra factor $v(q)$ in each vertex). The renormalized bubble
$({\bf ij})$  can be then obtained by summing up the series, consisting
from the bubbles $({\bf ij})$, connected by plasmon lines (see Fig.~1c).
The resulting expression for ${\bf P}_{\bf ij}(\omega)$ reads

\begin{eqnarray}\label{32}
{\bf P}_{\bf ij}(\omega)=
{P_{\bf ij}(\omega,)
\over 1-P_{\bf ij}(\omega)R_{\bf ij}(\omega)},
\end{eqnarray}
with

\begin{eqnarray}\label{33}
R_{\bf ij}(\omega)=\int {d{\bf q}\over (2\pi)^2}
|\langle {\bf i}|e^{i{\bf qr}}|{\bf j} \rangle|^2
v^2(q)\Pi(\omega, q).
\end{eqnarray}
Finally, replacing $P_{\bf ij}(\omega)$ in (\ref{31}) by
${\bf P}_{\bf ij}(\omega)$ we obtain

\begin{eqnarray}\label{34}
{\bf P}(\omega,q)={1\over A}
\sum_{\bf ij}|\langle {\bf i}|e^{i{\bf qr}}|{\bf j} \rangle|^2
{n_{\bf i}-n_{\bf j} \over \hbar\omega + E_{\bf i}-E_{\bf j} -
(n_{\bf i}-n_{\bf j})R_{\bf ij}(\omega)}.
\end{eqnarray}
The equations (\ref{30}), (\ref{33}), and (\ref{34}) form a closed
system which determines $\Pi (\omega, {\bf q})$ and, correspondingly, the
renormalized dispersion law of a plasmon in a self-consistent way. The
approximation made in order to get the closed system [replacement of a
block by a sought function  $\Pi (\omega, q)$] is known as the
effective-medium approximation.

The analysis of the system (\ref{30},\ref{33},\ref{34}) reveals that
the renormalization of the plasmon dispersion law is weak. Namely, the
appearence of the term $R_{\bf ij}(\omega)$ in the denominator of
(\ref{34}) has the physical meaning that a pair $({\bf ij})$ acquires
a finite life-time $\tau_{\bf ij}$ due to the interaction with
plasmons. We will show that this life-time is long, i.e.,
$1/\tau_{\bf ij}\ll \omega_{\bf ij}$. It can be readily seen that the
difference between the renormalized polarization
$\mbox{Re}{\bf P}(\omega,q)$ and $\mbox{Re}P(\omega,q)$ originates
from resonant pairs with $(\omega -\omega_{\bf ij})\sim 1/\tau_{\bf ij}$
and $r\sim \xi$ (note that the main term, $\mbox{Re}P(\omega,q)$, is
determined by the entire interval $\omega_{\bf ij}\sim \omega$). If we
neglect the dependence of the matrix element in (\ref{34}) on
$\omega_{\bf ij}$ then the renormalization correction to
$\mbox{Re}P(\omega,q)$ would be identically zero. A finite correction
results from a slight asymmetry of the matrix element within the
narrow interval  $(\omega -\omega_{\bf ij})\sim 1/\tau_{\bf ij}$. Then
the relative magnitude of the correction is of the order of
$1/\omega\tau_{\bf ij}$ with $\tau_{\bf ij}$ calculated for a pair
with $\omega_{\bf ij}=\omega$. Expecting $\tau_{\bf ij}$ to be long,
we can calculate it by substituting the nonrenormalized dispersion law
of a plasmon into Eq.~(\ref{33}). Performing the integration we obtain

\begin{eqnarray}\label{35}
R_{\bf ij}(\omega)={\hbar\over\tau_{\bf ij}}&\sim&\hbar\omega
\Biggl({\hbar\omega\over U}\Biggr)\Biggl({I_0\over U}\Biggr)^2
\ln^{-2}\Biggl({\hbar\omega\over I_0}\Biggr),~~~~~~~~~~\mbox{with gate,}
\nonumber\\
&\sim&\hbar\omega
\Biggl({\hbar\omega\over U}\Biggr)^3\Biggl({I_0\over U}\Biggr)^2
\ln^{-3}\Biggl({\hbar\omega\over I_0}\Biggr),~~~~~~~~\mbox{without gate.}
\end{eqnarray}
Since both ratios, $\hbar\omega/U$ and $I_0/U$ are small, the
correction to the dispersion law,
$\delta q(\omega)/q(\omega)\sim 1/\omega\tau_{\bf ij}$, is negligible.

In the next section we will see that the corresponding renormalization
of $\mbox{Re}\sigma (\omega)$ is much larger than the renormalization
of the dispersion law.

\section{Renormalization of the real part of the conductivity}
\label{sec5}

As we have seen in the previous section, the renormalization of the
polarization operator results in an appearance of
$i\hbar/\tau_{\bf ij}$ in the denominator of Eq.~(\ref{34}).
Correspondingly, the renormalized expression (\ref{4}) for
$\mbox{Re}\sigma(\omega)$ can now be rewritten as

\begin{eqnarray}\label{36}
\mbox{Re}\sigma(\omega)={e^2\omega\over \hbar A}\mbox{Im}
\sum_{\bf ij}\Biggl[{I_{ij}(x_i-x_j)\over \hbar\omega_{\bf ij}}\Biggr]^2
{1 \over \omega + \omega_{\bf ij} + i/\tau_{\bf ij}},
\end{eqnarray}
where the summation is performed over the singly occupied pairs
${\bf ij}$ and in this way the Coulomb correlations are taken into
account. In the limit $\tau_{\bf ij}\rightarrow\infty$ and for
$\omega >2I_0$ one should use the pair density $F(\Gamma,r)$, given by
Eq.~(\ref{20}), in order to perform the summation. The result is
determined by resonant pairs with
$\omega=\omega_{\bf ij}=\Gamma/\hbar$:

\begin{eqnarray}\label{37}
\mbox{Re}\sigma(\omega)={3\pi^2\over 4}
{e^2\over \hbar}(I_0g\xi^2)^2\Biggl(1+{U\over\hbar\omega}\Biggr)=
\sigma_0\Biggl(1+{U\over\hbar\omega}\Biggr).
\end{eqnarray}
We see that within the frequency interval $2I_0<\hbar\omega<U$ the
real part of the conductivity exceeds the Drude value due to Coulomb
correlations.

When calculating the correction to $\mbox{Re}\sigma(\omega)$ caused by
the finite value of $\tau_{\bf ij}$ it is important to realize that
$\hbar/\tau_{\bf ij}$ is maximal for soft pairs with small
$\omega_{\bf ij}$. This is because the matrix element
$\langle {\bf i}|e^{i{\bf qr}}|{\bf j} \rangle$ is proportional to
$1/\omega_{\bf ij}$. In the previous Section this was not important
since the correction to $\mbox{Re}P(\omega,q)$ came from the resonant
pairs only. Here, however, we have
$\mbox{Im}(\omega + \omega_{\bf ij} + i/\tau_{\bf ij})^{-1}
\propto 1/\tau_{\bf ij}\propto 1/\omega_{\bf ij}^2$, so that the soft
pairs give the main contribution to the correction
$\delta\mbox{Re}\sigma(\omega)$. Assuming $\omega_{\bf ij}\ll\omega$
we can present this correction in the form

\begin{eqnarray}\label{38}
\mbox{Re}\sigma(\omega)&=&{e^2\over \hbar\omega A}
\sum_{\bf ij}{[I_{ij}(x_i-x_j)]^2
\over (\hbar\omega_{\bf ij})^2\tau_{\bf ij}}
\nonumber\\
&=&{e^2\over \hbar\omega A}
\sum_{\bf ij}\Biggl[{I_{ij}(x_i-x_j)
\over \hbar\omega_{\bf ij}}\Biggr]^4
\int {d{\bf q}\over (2\pi)^2}
q^2v^2(q)\mbox{Im}\Pi(\omega, q),
\end{eqnarray}
where we have substituted $\hbar/\tau_{\bf ij}=R_{\bf ij}$ from
Eq.~(\ref{33}). The sum over pairs is again evaluated with the pair
density (\ref{20})

\begin{eqnarray}\label{39}
{1\over A}\sum_{\bf ij}\Biggl[{I_{ij}(x_i-x_j)
\over \hbar\omega_{\bf ij}}\Biggr]^4=
{1\over 4}\int d{\bf r}r^4I^4(r)
\int_{2I(r)}^{\infty}d\Gamma{F(\Gamma,r)\over\Gamma^4}.
\end{eqnarray}
The main contribution to the integral over $\Gamma$ comes from the
lower limit $\Gamma\sim I(r)$. Then the integral over $r$ is again
determined by $r\sim\xi$, so that the sum (\ref{39}) appears to be
$\sim I_0\xi^6g^2U$. As in the previous section, the value of the
integral in (\ref{39}) takes different values in the presence and in
the absence of a gate. Finally we obtain

\begin{eqnarray}\label{40}
{\delta\mbox{Re}\sigma(\omega)\over\sigma_0}
&\sim&
\Biggl({\hbar\omega\over U}\Biggr)^3\Biggl({\hbar\omega\over I_0}\Biggr)
\ln^{-2}\Biggl({\hbar\omega\over I_0}\Biggr),~~~~~~~~~~\mbox{with gate,}
\nonumber\\
&\sim&
\Biggl({\hbar\omega\over U}\Biggr)^5\Biggl({\hbar\omega\over I_0}\Biggr)
\ln^{-3}\Biggl({\hbar\omega\over I_0}\Biggr),~~~~~~~~\mbox{without gate.}
\end{eqnarray}
Comparing (\ref{35}) to (\ref{40}) we see that both corrections are of the
same order $\left(I_0/U\right)^2$ at $\hbar\omega\sim I_0$. However the
correction to ${\mbox Re}\sigma$ is much bigger at $\hbar\omega\gg I_0$.
This reveals a new   mechanism of absorption of a.c. field:
by resonant excitation of plasmons.
More precisely, the field polarizes the soft pairs (with
$\omega_{\bf ij}\sim I_0/\hbar$) and the induced polarization
excites the  plasmon waves. Therefore, the energy of a.c. field is effectively
absorbed by plasmons. The rapid increase of $\delta\sigma$ with $\omega$
is caused by the number (phase volume) of plasmons which absorb the field.

Note that with correction $\delta\sigma(\omega)$ the total
conductivity ${\mbox Re}\sigma(\omega)$
exhibits a rather complicated behavior. For $\hbar\omega<2I_0$ the
conductivity increases with $\omega$. Then it passes through a maximum
at $\omega\sim I_0/\hbar$ and falls off with $\omega$ according to
Eq.~(\ref{37}). However at $\hbar \omega\sim U(I_0/U)^{1/4}$ (with
gate) and  $\hbar \omega\sim U(I_0/U)^{1/6}$ (without gate) we have
$\delta{\mbox Re}\sigma(\omega)\sim {\mbox Re}\sigma(\omega)$
and the conductivity starts rising again. On the other hand, the
expression for $\delta{\mbox Re}\sigma$ was derived assuming that it
is small. Therefore in the region
$\delta{\mbox Re}\sigma >{\mbox Re}\sigma(\omega)$ the renormalization
of $\sigma(\omega)$ by plasmons is strong. In this case one cannot
calculate $\tau_{\bf ij}$ using the bare polarization operator. The
full analysis of the system (\ref{30},\ref{33},\ref{34}) in this
frequency range is out of the scope of the present paper.

\section{Conclusion}
\label{sec6}

In the present paper we argue that the wave of electric field can
propagate along the surface of the 2D Anderson insulator. The field
originates from the density fluctuations of localized electrons. One
should distinguish this wave from the usual plasmon in an ideal 2D gas
with the dispersion law given by Eq. (\ref{15}): ({\em i}) the
derivation of (\ref{15}) implies that $\omega\gg 1/\tau$ while we
predict the existence of a plasmon at much lower frequencies
$\omega\gtrsim I_0/\hbar$. The minimal frequencies $1/\tau$ and
$I_0/\hbar$ differ by a factor $\exp(\pi k_Fl)$; ({\em ii}) the
plasmon (\ref{15}) results from the solution of Eq. (\ref{13}) with
polarization operator calculated for free electrons. Within this
approximation Eq. (\ref{13}) has no solutions for {\em localized}
electrons. The solution appears only if one takes into account the
Coulomb correlations in the occupation numbers of the localized states.

The obvious consequence of the existence of a plasmon excitation is
that localized electrons ``feel'' a fluctuating electric field
${\cal E}(\omega,{\bf q})$ with the spectral density

\begin{eqnarray}\label{41}
\langle |{\cal E}(\omega,{\bf q})|^2\rangle=
{\kappa^2\over 2\pi e^2}{q^2v^2(q)\mbox{Im}{\bf P}(\omega,q)\over
[1-v(q)\mbox{Re}{\bf P}(\omega,q)]^2+[v(q)\mbox{Im}{\bf P}(\omega,q)]^2},
\end{eqnarray}
where ${\bf P}(\omega,q)$ is the polarization operator (\ref{34}). The
spectral density (\ref{41}) has a peak at $\omega=\omega(q)$
corresponding to the dispersion law of a plasmon, which is different
with and without gate.

The basic assumption of our theory is that in the presence of
interactions the Anderson insulator is characterized by two energy
scales: $I_0$ - the spacing between energy levels in the area $\xi^2$
and the charging energy $U$ which is the Coulomb interaction of two
localized electrons separated by a distance $\sim\xi$. This energy
modifies the pair density $F(\omega,r)$. Such a modification results
in an enhancement\cite{shk81,efr81,efr85,efr85b} of the dissipative
conductivity
$\mbox{Re}\sigma(\omega)$. If we denote as $\rho(\omega)$ the matrix
element of $r$ calculated for a pair with frequency $\omega$ then

\begin{eqnarray}\label{42}
\mbox{Re}\sigma(\omega)\propto \omega\rho^2(\omega)F(\omega, r_{\omega}).
\end{eqnarray}
As it was discussed in the Introduction, it is plausible to assume
that for the Anderson insulator in the absence of interactions
$\rho(\omega)$ behaves as
$\rho(\omega)\sim r_{\omega}$ for $\omega\lesssim I_0/\hbar$ and
$\rho(\omega)\sim\xi I_0/\omega$ for $\omega\gtrsim I_0/\hbar$. Since
the pair density in the absence of interactions is proportional to
$\omega$ we reproduce Eqs.~(\ref{11}) and (\ref{12}):

\begin{mathletters}\label{43}
\begin{eqnarray}
\mbox{Re}\sigma(\omega)&\propto&\omega^2,
{}~~~~~~~~~~~~\mbox{for}~~~\omega\lesssim I_0/\hbar,\label{43a}\\
\mbox{Re}\sigma(\omega)&=& const,
{}~~~~~~~~~\mbox{for}~~~\omega\gtrsim I_0/\hbar,\label{43b}
\end{eqnarray}
\end{mathletters}
Where we neglected a logarithmic factor in (\ref{43a}).
We have assumed that the frequency dependence of $\rho(\omega)$
remains the same in the presence of interactions.\cite{note} At the
same time, $F(\omega,\xi)$ is changed drastically by interactions:
$F(\omega,\xi)=const$ for $\omega\lesssim U/\hbar$ and
$F(\omega,\xi)\propto\omega$ for $\omega\gtrsim U/\hbar$. As a result,
we get the following frequency dependence of $\mbox{Re}\sigma(\omega)$
in the presence of interactions [see also (\ref{37})]:

\begin{mathletters}\label{44}
\begin{eqnarray}
\mbox{Re}\sigma(\omega)&\propto& \omega,
{}~~~~~~~~~~~~~~\mbox{for}~~~\omega\lesssim I_0/\hbar,\label{44a}\\
\mbox{Re}\sigma(\omega)&\propto& 1/\omega,
{}~~~~~~~~~~~\mbox{for}~~~I_0/\hbar\lesssim\omega\lesssim U/\hbar,\label{44b}\\
\mbox{Re}\sigma(\omega)&=& \mbox{const},
{}~~~~~~~~~\mbox{for}~~~\omega\gtrsim U/\hbar,\label{44c}
\end{eqnarray}
\end{mathletters}
This simple analysis shows that $\mbox{Re}\sigma(\omega)$ exhibits
a maximum at $\omega\sim I_0/\hbar$ (see also the end of the previous
section). In fact, this maximum is
intimately related to the existence of a plasmon. Indeed, for
Eq. (\ref{13}) to have solutions we need a positive sign of
$\mbox{Re}{\bf P}(\omega,q)$, which is equivalent to a positive
$\mbox{Im}\sigma(\omega)$. At the same time, $\mbox{Im}\sigma(\omega)$
can be obtained from $\mbox{Re}\sigma(\omega)$ using the
Kramers-Kronig relation. In order to trace how the modification of
$\mbox{Re}\sigma(\omega)$ by Coulomb correlations leads to the change
of sign of $\mbox{Im}\sigma(\omega)$, we can interpolate the
frequency dependence of $\rho(\omega)$ as
$\rho(\omega)\propto [(\hbar\omega)^2+4I_0^2]^{-1/2}$ and the
frequency dependence of $F(\omega,\xi)$ as
$F(\omega,\xi)\propto (\hbar\omega + U)$. Then we have

\begin{eqnarray}\label{45}
\mbox{Re}G(\omega)={\hbar\omega(\hbar\omega+U)
\over (\hbar\omega)^2+4I_0^2},~~~~~~~~~~\omega>0,
\end{eqnarray}
where $G={\sigma\over e^2/\hbar}$ is the conductance. It can be easily
seen that Eq. (\ref{45}) reproduces correctly all the limiting
cases (\ref{44}) and (\ref{44}) both with and without charging effect.
The imaginary
part of the conductance calculated from (\ref{45}) has a simple form:

\begin{eqnarray}\label{46}
\mbox{Im}G(\omega)=
-{\hbar\omega[2I_0-(2U/\pi)\ln(\hbar\omega/ 2I_0)]
\over (\hbar\omega)^2+4I_0^2},~~~~~~~~~~\omega>0,
\end{eqnarray}
In Fig. 2 we have plotted
$\mbox{Re}G(\omega)$ and $\mbox{Im}G(\omega)$ for different ratios
$U/2I_0$. We see that in the absence of Coulomb correlations ($U=0$),
$\mbox{Im}G(\omega)$ is strictly negative (in our calculation in
Section \ref{sec2} it turns to zero for $\omega>I_0/\hbar$). However,
at finite $U$ the change of sign occurs at
$\omega=(2I_0/\hbar)\exp(\pi I_0/U)$. For $U\gg I_0$ this frequency is
just $2I_0/\hbar$.

As a final remark, let us outline the difference between our approach
and that of Ref.~\onlinecite{fle78}. In Ref.~\onlinecite{fle78} the
authors addressed electron-electron interactions in the strongly
localized regime when the localization radius is much smaller than the
interpair separation. In this case singly-occupied pairs can be
considered as point-like dipoles. The dipole moment ${\bf p}_k$
induced by an external field ${\cal E}_0e^{-i\omega t}$ can be written
as

\begin{eqnarray}\label{47}
p_k^{\mu}=\alpha_k^{\mu\nu}[{\cal E}_0^{\nu}+
\sum_{l\neq k}{\cal E}_k^{(l)\nu}],
\end{eqnarray}
where ${\cal E}_k^{(l)\nu}$ is a component $\nu$ of the electric field
caused by polarization of a dipole $l$ which acts on the dipole $k$
(summation over repeating indices $\nu$ is implied). The
polarizability $\alpha_k^{\mu\nu}$ of a dipole $k$ has the form

\begin{eqnarray}\label{48}
\alpha_k^{\mu\nu}={2e^2\over\hbar}{\rho_k^{\mu}\rho_k^{\nu}\omega_k
\over \omega^2-\omega_k^2}
={e^2\over\hbar}\rho_k^{\mu}\rho_k^{\nu}\sum_{\bf ij}P_{\bf ij},
\end{eqnarray}
where $\rho_k^{\mu}$ and $\omega_k$ are correspondingly the matrix
element and the frequency of the dipole $k$. Here $P_{\bf ij}$ is
given by Eq.~(\ref{31}) with the states ${\bf i}$ and ${\bf j}$ making
up the dipole $k$. To keep the discussion
simple we will assume that the polarizability is isotropic, i.e.,
$\alpha_k^{\mu\nu}=\alpha_k\delta_{\mu\nu}$. The field
${\cal E}_k^{(l)\nu}$ can, in turn, be expressed through ${\bf p}_k$
as

\begin{eqnarray}\label{49}
{\cal E}_k^{(l)\mu}=-{\partial\over\partial R^{\mu}}
{{\bf p}_l{\bf R}\over R^3}\Biggl|_{{\bf R}={\bf R}_k-{\bf R}_l},
\end{eqnarray}
where ${\bf R}_k$ is the position of the dipole $k$.

Upon substituting (\ref{49}) into (\ref{47}) we obtain an infinite
system of linear equations. It can be easily shown that iterating this
system leads  to the renormalization of polarizabilities of dipoles
$\alpha_k$

\begin{mathletters}\label{50}
\begin{eqnarray}
\alpha_k={e^2\over\hbar}\rho_k^2
\sum_{\bf ij}{P_{\bf ij}\over 1-P_{\bf ij}\Sigma_k}
={2e^2\over\hbar}{\rho_k^2(\omega_k+\Sigma_k)
\over \omega^2-(\omega_k+\Sigma_k)^2},\\
\Sigma_k=-\sum_l \alpha_l
\Biggl({\partial\over\partial R^{\mu}}{R^{\nu}\over R^3}\Biggr)
\Biggl({\partial\over\partial R^{\nu}}{R^{\mu}\over R^3}\Biggr)
\Biggl|_{{\bf R}={\bf R}_k-{\bf R}_l}+\cdots.
\end{eqnarray}
\end{mathletters}
Here $\Sigma_k$ is the self-energy. Fleishman and Anderson\cite{fle78}
anylized this self-energy using the arguments similar to those put
forward by Anderson\cite{and58} when he demonstrated the existence of
localization transition for eigenfunctions of the Schroedinger
equation with disorder. They argued that $\mbox{Im}\Sigma$ takes a
finite value with non-zero probability. This means that in the absence
of an external field the system

\begin{eqnarray}\label{51}
{\bf p}_k+\alpha_k\nabla\sum_{l\neq k}
{{\bf p}_l{\bf R}\over R^3}\Biggl|_{{\bf R}={\bf R}_k-{\bf R}_l}=0,
\end{eqnarray}
has delocalized solutions.\cite{lev90} In other words, by analogy to the
Schoedinger equation in the tight-binding approximation, the
eigenstates $\{{\bf p}_k\}$ of the system ({\ref{51}) extend throughout
the entire volume.

Fleishman and Anderson\cite{fle78} considered a three-dimensional
system and neglected the Coulomb correlations. Note that if one
rewrites the system (\ref{51}) in the momentum representation and
averages it by factorizing the average of the product $\alpha{\bf p}$
(mean-field) then one arrives at the plasmon equation (\ref{13}). As it was
demonstrated in Section \ref{sec2}, this equation has no propagating
solutions in the absence of Coulomb correlations. Our central point is
that {\em with} the Coulomb correlations the delocalized solution
exists even at the mean-field level. This solution, that is surface
plasmon, is specific for the two-dimensional system and is
characterized by the dispersion law $\omega(q)$.

An interesting question that could be addressed
within the same approach is how the Coulomb correlations modify the
a.c. Hall conductivity $\sigma_{xy}(\omega)$ of the Anderson
insulator.\cite{imr93,zha92}

\acknowledgments
The authors are indebted to B.~I. Shklovskii for numerous discussions.
In fact, Sections \ref{sec4}-\ref{sec6} have emerged from his hot contesting
of our basic assumptions.

\begin{figure}\label{fig1}
\caption{(a) Diagrammatic presentation of the density-density correlation
function $\Pi(\omega,q)$; bubble $({\bf ij})$ stands for
$P_{\bf ij}(\omega)$ while dashed line corresponds to the Coulomb
interaction $v(q)$. Each vertex contains matrix element
$\langle {\bf i}|e^{i{\bf qr}}|{\bf j} \rangle$.
(b) The block connecting two bubbles $({\bf ij})$ is replaced by a
plasmon line. (c) The renormalization of the bubble $({\bf ij})$
caused by interaction with plasmons.}
\end{figure}

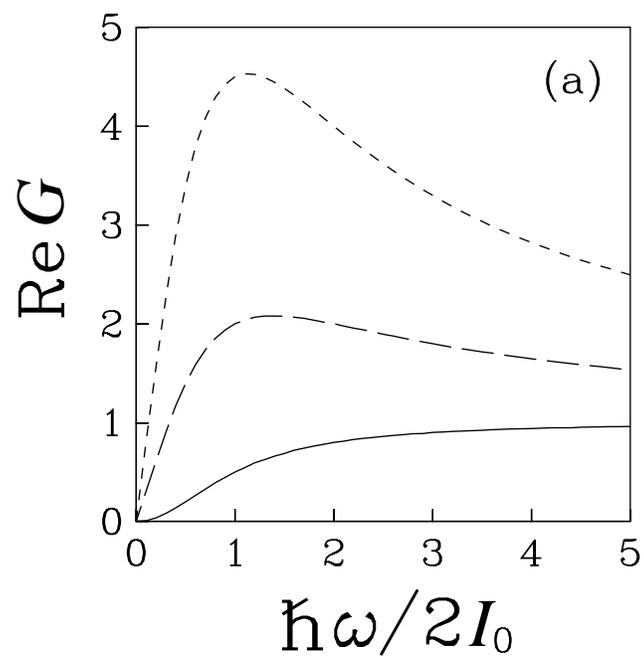
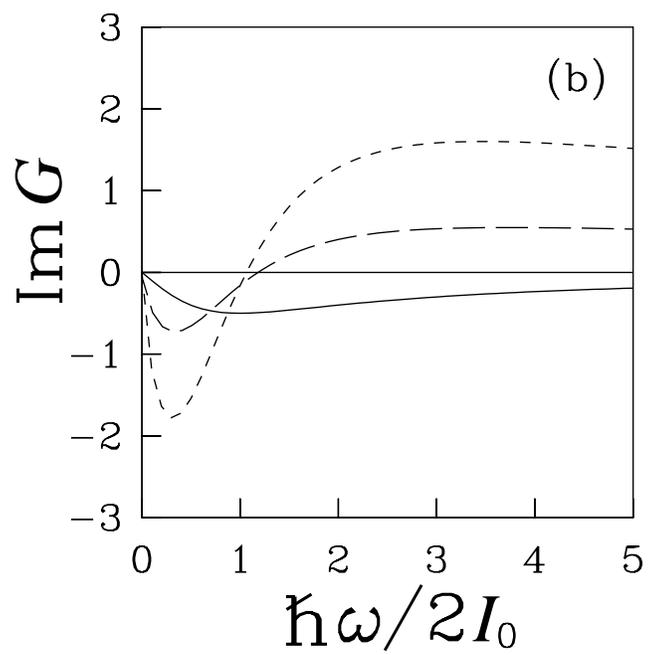
\begin{figure}\label{fig2}
\caption{The real (a) and imaginary (b) parts of conductance is plotted as a
function of dimensionless frequency $\hbar\omega/2I_0$ for
$U/2I_0=0$ (solid line), $U/2I_0=3$ (long-dashed line), and
$U/2I_0=8$ (dashed line).}
\end{figure}

\begin{references}
\bibitem{note} Note that in Ref. \onlinecite{pol93} it was assumed
that interactions result in the enhancement of $\rho(\omega)$ by a
factor $U/I_0$.
\end{references}
\end{document}